\documentclass[preprint]{aa} 

\usepackage{epsfig}
\usepackage{graphicx}
\usepackage{txfonts}
\usepackage[normalem]{ulem}

\newcommand{\outvect}{\mathbf{x}}
\newcommand{\kvect}{\mathbf{y}}
\newcommand{\knormvect}{\mathbf{k}}

\begin{document}

\title{Kernel-nulling for a robust direct interferometric \\ detection of extrasolar planets}

\author{
  Frantz Martinache\inst{1} \&
  Michael J. Ireland\inst{2}.
}

\institute{Laboratoire Lagrange, Université Côte d’Azur, Observatoire de la
  Côte d’Azur, CNRS, Parc Valrose, Bât. H. FIZEAU, 06108 Nice, France
  \email{frantz.martinache@oca.eu} \and
  Research School of Astronomy \& Astrophysics, Australian National University,
  Canberra ACT 2611, Australia }


\abstract
{Combining the resolving power of long-baseline interferometry with the
  high-dynamic range capability of nulling still remains the only technique
  that can directly sense the presence of structures in the innermost regions
  of extrasolar planetary systems.}
{Ultimately, the performance of any nuller architecture is constrained by the
  partial resolution of the on-axis star whose light it attempts to cancel
  out. However from the ground, the effective performance of nulling is
  dominated by residual time-varying instrumental phase  and background
    errors that keep the instrument off the null. Our work investigates
    robustness against instrumental phase.}
{We introduce a modified nuller architecture that enables the extraction of
  information that is robust against piston excursions. Our method generalizes
  the concept of kernel, now applied to the outputs of the modified nuller so
  as to make them robust to second order pupil phase error. We present the
  general method to determine these kernel-outputs and highlight the benefits
  of this novel approach.}
{We present the properties of VIKiNG: the VLTI Infrared Kernel NullinG, an
  instrument concept within the Hi-5 framework 
  for the 4-UT VLTI infrastructure that takes advantage of
  the proposed architecture, to produce three self-calibrating nulled outputs.}
{Stabilized by a fringe-tracker that would bring piston-excursions down to 50
  nm, this instrument would be able to directly detect more than a
  dozen extrasolar planets so-far detected by radial velocity only, as well as
  many hot transiting planets and a significant number of very young exoplanets.}


\titlerunning{Kernel-nulling interferometry}

\maketitle


\section{Introduction}
\label{sec:intro}

The direct imaging of extrasolar planets from the ground remains an incredibly
challenging objective that requires the simultaneous combination of high angular
resolving power, required to see objects separated by a few astronomical units
and located tens of parsecs away, with high-dynamic imaging capability to
overcome the large contrast between the faint planet and its bright host
star. This objective is doubly limited by the phenomenon of diffraction, that
sets a limit to the resolving power of a telescope or interferometer, and
produces diffraction features such as rings, spikes, fringes and speckles whose
contribution to the data dominates that of the faint structures one attempts to
detect, by several orders of magnitude.

A high-contrast imaging device, be it a coronagraph \citep{1932ZA......5...73L}
when observing with a single telescope or a nuller \citep{1978Natur.274..780B}
when using an interferometer, is a contraption devised to attenuate the static
diffraction-induced signature of one bright object in the field, while
transmitting the rest of the field. Very elegant and effective solutions have
been devised \citep{2003A&A...404..379G, 2005ApJ...618L.161S,
  2010ApJ...709...53M}, that can theoretically deliver data where the
contribution of the bright star is attenuated to up to ten orders of magnitudes
\citep{2007Natur.446..771T} and a few such coronagraphs are currently in
operation on ground based observing facilities. Their high-contrast imaging
capability is however severely affected by the less than ideal conditions they
experience when observing through the atmosphere, even
\citep{2004ApJ...612L..85A} with correction provided by state-of-the-art
extreme adaptive optics (XAO) systems like VLT/SPHERE
\citep{2006Msngr.125...29B}, the Gemini Planet Imager
\citep{2014PNAS..11112661M} or the Subaru Telescope SCExAO
\citep{2015PASP..127..890J}.

The position of an aberration-induced speckle in the field is related to a
sinusoidal wavefront modulation across the aperture of the instrument, and the
contrast $c$ of this speckle at wavelength $\lambda$ is directly related to the
amplitude $a$ of the modulation, using the following simple relation:

\begin{equation}
  c = \left(\frac{2\pi a}{\lambda}\right)^2,
\end{equation}

\noindent
which can be used to estimate how to translate a raw-contrast objective into a
requirement on the wavefront stability. Thus, regardless of the architecture of
the high-contast device, a raw contrast $c=10^{-6}$ ambition for an instrument
observing in the H-band ($\lambda = 1.6\, \mu$m) translates into a wavefront
quality requirement better than 0.25 nm, which is more than two orders of
magnitude beyond what state of the art XAO systems are able to deliver
\citep{2016JATIS...2b5003S}.

Reported recent detections of extrasolar planet companions
\citep{2015Sci...350...64M, 2017A&A...605L...9C}, owe much to post-processing
techniques such as angular differential imaging \citep{2006ApJ...641..556M}
that make it possible to disantangle genuine structures present in the image
from residual diffraction features \citep{2008Sci...322.1348M} that otherwise
dominate it. To increase the impact of the high-contrast device in the
pre-processing stage, one approach might be to look into solutions that do not
necessarily produce the highest performance when operating in ideal but rarely
occuring observing conditions, but instead integrate some form of robustness
against small perturbations. The work described in this paper is a step
in this direction.

An alternative observing technique to XAO-fed coronagraphy for high-contrast
detection of extrasolar planets is to use long-baseline nulling-interferometry.
Thanks to their higher angular resolution, long-baseline nulling
interferometers allow the observation of planets much closer to the star than
coronagraphs or to use a longer mid-infrared wavelength, where the expected
star-planet contrast is expected to be more favorable
\citep{2005ApJ...626..523C}. Very much like for ground-based coronagraphy, the
effective actual high-contrast detection potential of nulling is constrained by
variable observing conditions, that result in fluctuations of the thermal
background as well as small piston excursions, minimized by fringe tracking,
that keep the observation off the null \citep{2012ApJ...748...55S}. For
instance, N-band nulling instruments such as the Keck Interferometric Nuller
(KIN) and the Large Binocular Telescope Interferometer (LBTI) are limited to
constrasts of a few $10^{-4}$ to a few $10^{-3}$ by residual background errors
(e.g., \citet{Colavita2009,Defr2016}, while at shorter wavelength the Palomar
Fiber Nuller (PFN) was limited to contrast of a few $10^{-4}$ due to
high-frequency residual phase errors (Mennesson et al. 2011). Here too,
post-acquisition analysis of the distribution of the measured null
\citep{2011ApJ...729..110H, 2011ApJ...736...14M} make it possible to further
characterize the true null depth and improve the contrast detection limits, an
approach refered to as Null Self-Calibration (NSC). This approach requires a
nuller to detect off-null light with high signal-to-noise within an
instrumental coherence time, so is not applicable to observations anywhere near
the shot-noise limit of a nulling instrument. It is also currently not
applicable to array configurations with more than two telescopes.

Instead, and similarly to high-contrast imaging, pre-processing techniques can
be used to improve the null depth and its robustness against
perturbations. Over the years, the original idea of \citet{1978Natur.274..780B}
has been refined to improve the rejection of the nuller, usually by
simultaneously combining more than two apertures \citep{1997ApJ...475..373A}
and optimizing the internal structure of the nuller
\citep{2013PASP..125..951G}. However, a major limiting factor in exploring
these multi-aperture designs has been the difficulty in creating optical
devices of sufficient precision and complexity. One avenue which has shown
rapid recent process is mid-infrared photonic beam combination, both in
ultrafast laser inscription lithography in chalcogenide
\citep{2017A&A...602A..66T} and fluoride \citep{2015OExpr..23.7946G}
substrates, and in planar photolithography based devises using chalcogenide
glass \citep{2017OExpr..25.3038K} and lithium niobate
\citep{2009OExpr..1718489H,2014SPIE.9146E..2IM}. These emerging technological
platforms are in need of clear required performance metrics and baseline
architectures to define succesful technological development for astrophysics.

In this paper, we present a true self-calibration technique, more akin to the
properties of observable quantities like closure-phase
\citep{1958MNRAS.118..276J}, which takes advantage of the coupling between
atmospheric induced piston errors along baselines forming a triangle, to
produce from a finite set of polluted raw phase measurements, a subset of clean
observable quantities, robust against residual piston errors. Shown to be
usable in the optical regime \citep{1986Natur.320..595B}, it is extensively
used during non-redundant aperture masking interferometry observations
\citep{2000PASP..112..555T} and also takes advantage of the correction provided
by AO \citep{2006SPIE.6272E.103T}, as it enables long exposure observations
with improved sensitivity. Using closure-phase, VLTI/PIONIER observations
  achieve contrast detection limits of a few $10^{-3}$ \citep{Absil2011} alone,
  without a nuller. The notion of closure-phase was later shown to be a
special case of kernel-phase \citep{2010ApJ...724..464M}: instead of looking
for closure triangles in an aperture, one treats the properties of an
interferometer globally, using a single linear operator $\mathbf{A}$ to
describe the way instrumental phase propagates in the relevant observable
parameter space (the Fourier-phase, in the case of kernel-phase), and looks for
linear combinations of polluted data that reside in a space orthogonal to the
source of perturbation, described by the row-space of $\mathbf{A}$.

This paper describes how the design of a nuller can be modified to take the
possibility of self-calibration into account, to produce observable quantities
that are robust against second-order atmospheric-piston-induced phase
excursions. The paper uses a generic recipe that is applied to a four-beam
nulling combiner, which is the most relevant case for exploiting the
capabilities of the existing Very Large Telescope Interferometer (VLTI), within
the framework recently provided by the Hi-5 project
\citep{2018arXiv180104148D}.

\section{Enabling self-calibration for a nuller}
\label{sec:enbl}

\subsection{Nuller design and parametrisation}
\label{sec:design}

The nuller we are looking at is a combiner taking four inputs of identical
collecting power and designed to produce one bright output and three dark
ones. This design ignores the true location of the sub-apertures making up the
interferometric array, and how these can impact the order of the null
\citep{2013PASP..125..951G}.

Such a four-beam nuller can be represented by a 4x4 matrix $\mathbf{N}$, acting
on the four input complex amplitudes collected by the four apertures, and
producing the expected outputs. For the nuller we consider here:

\begin{equation}
  \mathbf{N} = \frac{1}{\sqrt{4}} \times
  \begin{bmatrix}
    1 & 1 & 1 & 1 \\
    1 & 1 & -1 & -1 \\
    1 & -1 & 1 & -1 \\
    1 & -1 & -1 & 1
    \end{bmatrix}.
\end{equation}

Except for the first row of this matrix for which the input complex amplitudes
are constructively combined, each row combines differences of complex
amplitudes that would result, for a single on-axis unresolved source and in the
absence of atmospheric piston, in a dark output. The global $1/\sqrt{4}$ (=0.5)
coefficient makes $\mathbf{N}$ a complex unitary matrix, accounting for the
fact that the interferometric recombination process preserves total flux:
$||\mathbf{N} \cdot x ||^2 = || x ||^2$. We have also considered a 4x4 matrix
that is constructed from two 2x2 nullers, where the bright outputs are
hierarchically combined in a second 2x2 nuller. The result from that
architecture is less symmetrical, but is not qualitatively different from the
results presented here.

Whereas the raw interferometric phase per baseline is linearly related to the
instrumental phase, making the definition of closure- and kernel-phase
reasonably direct, the output of a nuller is a quadratic function of piston
excursions \citep{2012ApJ...748...55S}. Of the four sub-apertures, one, labeled
$T_0$ is chosen as a phase reference so that phase or piston values are quoted
relative to this sub-aperture. The remaining degrees of freedom form a
three-parameter (correlated) piston vector $p$ that translates into the 
chromatic phase
$\varphi = 2 \pi p / \lambda$. Assuming that the source is unresolved by the
interferometer, a first order Taylor expansion of piston dependance of the
input electric field simply writes as:

\begin{equation}
  E_k = \exp{(-j\varphi_k)} \approx 1 - j \varphi_k.
  \label{eq:approx}
\end{equation}

Plugging these electric field as inputs to the nulling matrix $\mathbf{N}$, one
can write the equations for the three nulled intensities, valid to second order
in input phase:

\begin{equation}
  \outvect = \frac{1}{4} \times
  \begin{bmatrix}
    (\varphi_1 - \varphi_2 - \varphi_3)^2 \\
    (-\varphi_1 + \varphi_2 - \varphi_3)^2 \\
    (-\varphi_1 - \varphi_2 + \varphi_3)^2
  \end{bmatrix}.
  \label{eq:i1}
\end{equation}

Further expansion shows that the piston induced leak of the nuller is a
function of six parameters: three second order terms $(\varphi_k)^2$ and three
crossed-terms $\varphi_k \times \varphi_l$. With only the three relations
summarized by Equation \ref{eq:i1}, the problem is underconstrained and does
not permit the building of a set of kernels. To build kernels from the 
output of a
combiner, one needs to further break down each nuller output into two
non-symmetric outputs that will help discriminate variations in the two parts
of the complex visibilities, when properly mixed. This split-and-mix operation
can be represented by the following complex linear operator $\mathbf{S}$ that
enables the proper sensing of the nuller output:


%
%

\begin{equation}
  \mathbf{S} = \frac{1}{\sqrt{4}} \times
  \begin{bmatrix}
    1 & e^{i\theta} & 0  \\
    -e^{-i\theta} & 1 & 0  \\

    1 & 0 &  e^{i\theta}\\
    -e^{-i\theta} & 0 & 1\\
    
    0 & 1 & e^{i\theta} \\
    0 & -e^{-i\theta} & 1
    \end{bmatrix},
\end{equation}

\noindent
where $\theta$ is a pre-defined phase offset and $1/\sqrt{4}$ (= 0.5) a factor
that accounts for the total flux preservation when splitting each nulled output
into four. A detector placed downstream of this final function records a
now six-component intensity vector $\outvect$ recording the square modulus
associated to each output.

\begin{figure}
  \includegraphics[width=\columnwidth]{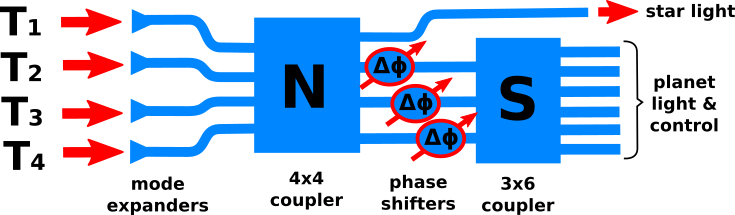}
  \caption{Schematic representation of the proposed two-stage nuller
    architecture. The first 4x4 coupler stage implements the nulling function
    described by the matrix $\mathbf{N}$ introduced in Section
    \ref{sec:design}. The second 3x6 coupler implements the sensing function
    described by the matrix $\mathbf{S}$. In between the two stages, modulated
    phase shifters are inserted so as to eliminate background fluctuations.}
  \label{f:concept_modulation}
\end{figure}

A practical implementation of a nuller has to deal with not only residual
starlight and phase-noise, but also fluctuating backgrounds and detector
noise. This means that a temporal modulation function is also required in
addition to the nulling function. Figure~\ref{f:concept_modulation} shows a
schematic representation of a possible interface between the two functions. By
modulating the phase shifters, the 6 nulled outputs can be rapidly permuted,
enabling the final signal to be obtained from synchronously demodulated
outputs. In addition, for faint targets, the starlight may not clearly be
detectable above a variable thermal background, meaning that even the star
light channel may need to be modulated, in order to apply the correct
normalisation to the planet light outputs. In any case, maintaining long-term
amplitude balance between the inputs requires either modulation or independent
photometric channels.

The concept described in the rest of the paper will ignore these background
fluctuations considerations and the modulation that would otherwise be required
to account for it: the nulling and sensing functions can therefore be combined
into a single six-by-four operator $\mathbf{M}$ that takes the four input
complex amplitudes incoming from the four telescopes and produces six nulled
output complex amplitudes:

\begin{equation}
  \mathbf{M} = \frac{1}{4} \times
  \begin{bmatrix}

    1 + e^{i\theta}  & 1 - e^{i\theta}   & -1 + e^{i\theta}  & -1 - e^{i\theta} \\
    1 - e^{-i\theta} & -1 - e^{-i\theta} &  1 + e^{-i\theta} & -1 + e^{-i\theta} \\
    1 + e^{i\theta}  & 1 - e^{i\theta}   & -1 - e^{i\theta}  & -1 + e^{i\theta}  \\
    1 - e^{-i\theta} & -1 - e^{-i\theta} & -1 + e^{-i\theta}  & 1 + e^{-i\theta} \\
    1 + e^{i\theta}  & -1 - e^{i\theta}  & 1 - e^{i\theta} & -1 + e^{i\theta}   \\
    1 - e^{-i\theta} & -1 + e^{-i\theta} & -1 - e^{-i\theta} & 1 + e^{-i\theta}
    \end{bmatrix}.
\end{equation}

\begin{figure}
  \includegraphics[width=\columnwidth]{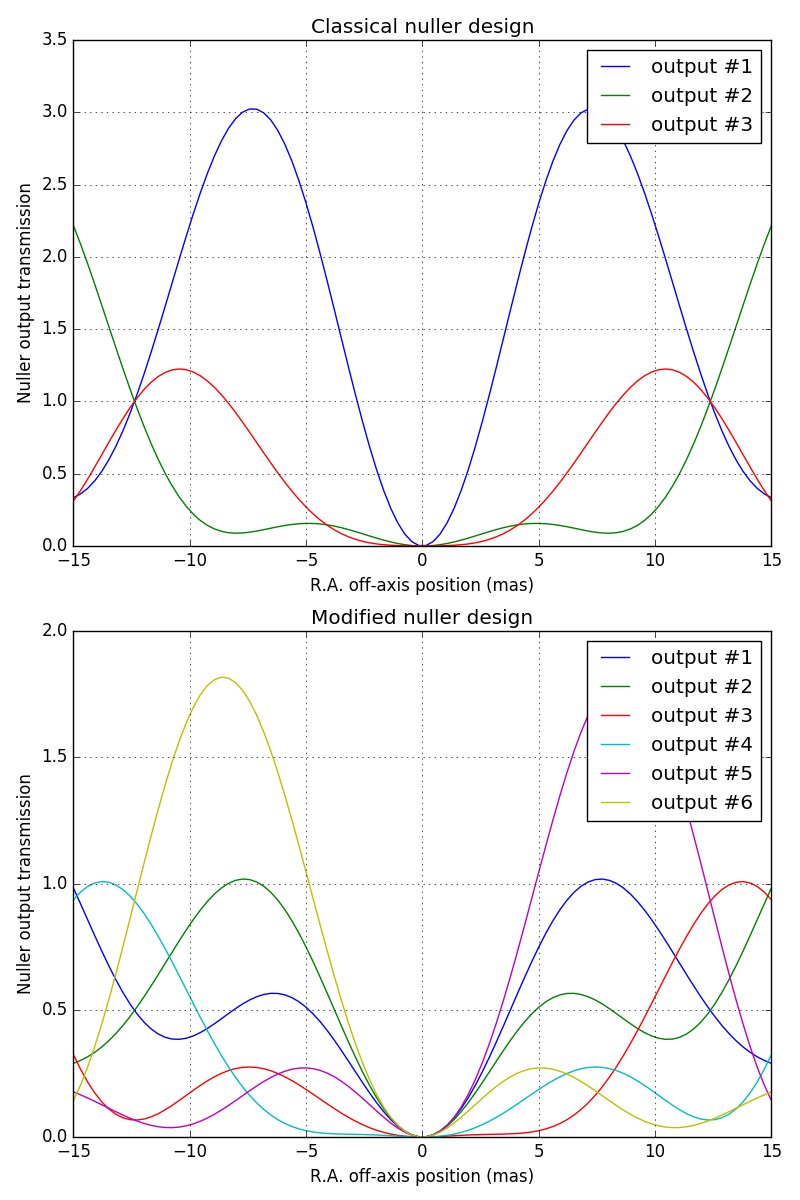}
  \caption{Comparison of the outputs of two nuller architectures as a function
    of right-ascension (R.A.) offset in milli-arcsecond (mas) for an in-line
    non-redundant array (aperture coordinates listed in Table \ref{tbl:line}):
    the classical design, corresponding to the three nulled outputs of the
    matrix $\mathbf{N}$ is at the top and the modified design, incorporating
    the mixing function described in the main text of the paper is at the
    bottom. In both cases, the output unit is in multiples of the transmission
    of a single telescope.}
  \label{f:nuller_out_plots}
\end{figure}

A detector placed downstream of the combiner now records a six-component
intensity vector $\outvect = || \mathbf{M} \cdot \mathbf{E}||^2$. To compare
the properties of this modified nuller design to those of the classical one,
Figure \ref{f:nuller_out_plots} presents a series of transmission curves of the
two nullers for an in-line non-redundant array of coordinates listed in Table
\ref{tbl:line}, and observing in the L-band ($\lambda = 3.6\,\mu$m), as a
function of source position offset relative to the null. The phase shifting
parameter of the mixing function will from now on be set to $\theta = \pi/2$,
as this specific value allows to write all matrices explicitly.

\begin{table}
  \caption{East and North coordinates (in meters) for a fictive non-redundant
    linear array, used in Sections \ref{sec:design} and \ref{sec:knulling} to
    illustrate some properties of the proposed nuller architecture.}
  \label{tbl:line}
  \centering
  \begin{tabular}{ l r r}
    \hline\hline
    Station & E & N \\
    \hline
    T1      & 0.0  & 0.00 \\
    T2      & 10.0 & 0.00 \\
    T3      & 40.0 & 0.00 \\
    T4      & 60.0 & 0.00 \\
    \hline
\end{tabular}
\end{table}

On-axis, the proposed architecture still behaves like a nuller with zero
transmission when operating in perfect conditions. Besides the expected
multiplication of outputs going from the classical to the modified nuller
design, a major difference lies in the symmetry properties of the outputs:
whereas the classical nuller features response curves that are symmetric
relative to the on-axis reference, the modified nuller outputs are
anti-symmetric and therefore allow to discriminate a positive from a negative
offset position, and give a stronger constraint on the position of a companion
around a bright star, from a single observation.

\subsection{Kernel-nulling}
\label{sec:knulling}

The motivation for the proposed architecture is the ability to build from the
six outputs of the combined for each acquisition, a sub-set of observable
quantities that exhibit some further robustness against residual piston
errors. In a classical (ie. non-nulling) combiner, the four input beam
interferometer gives access to up to six distinct baselines that can produce up
to three-closure phases \citep{2000plbs.conf..203M}, so one expects a
satisfactoy nuller architecture should produce three kernels on a non-redundant
array.

With one of the four sub-apertures chosen as zero-reference for the phase, the
aperture phase of a coherent point-like source reduces to a three-component
vector $\varphi$. When everything is in phase ($\varphi = 0$), the system sits
on the null, where the first order derivative terms of both phase and amplitude are all zeros (see the
bottom panel of Fig. \ref{f:nuller_out_plots}). Piston-induced leaked intensity
$\Delta\outvect$ by the nuller will therefore be dominated by second order
terms, whose impact can be estimated by measuring the local curvature. With
three degrees of freedom, six second order terms need to be accounted for:
three second-order partial derivatives and three second-order mixed
derivatives.

The response of the six intensity outputs to these six second-order
perturbations is recorded in a $6 \times 6$ matrix $\mathbf{A}$, analoguous to
the phase transfer matrix introduced by \citet{2010ApJ...724..464M} to find the
kernel of the information contained in the Fourier-phase, but generalized to
encode the impact of second-order differences in the pupil plane phase vector
on the output of a nuller:

\begin{equation}
  \Delta\outvect = \mathbf{A} \cdot
  \left[
    \frac{\partial^2\outvect}{\partial\varphi_1^2},
    \frac{\partial^2\outvect}{\partial\varphi_2^2},
    \frac{\partial^2\outvect}{\partial\varphi_3^2},
    \frac{\partial^2\outvect}{\partial\varphi_1\partial\varphi_2},
    \frac{\partial^2\outvect}{\partial\varphi_1\partial\varphi_3},
    \frac{\partial^2\outvect}{\partial\varphi_2\partial\varphi_3}
  \right]^T.
\end{equation}

Just like in the case of kernel-phase, depending on the properties of
$\mathbf{A}$, it may be possible to identify a sub-set of linear combinations
of rows of $\mathbf{A}$ which combined into an new kernel operator
$\mathbf{K}$, will verify:

\begin{equation}
  \mathbf{K} \cdot \mathbf{A} = \mathbf{0}.
\end{equation}

When the same kernel operator is applied to the raw output vector $\outvect$
of the nuller, it results in a smaller set of observable quantities:
$\mathbf{K} \cdot \outvect$ which are independent of second-order phase
differences in the pupil plane.

One of the most robust ways to produce the kernel operator is to compute the
singular value decomposition (SVD) of $\mathbf{A} =
\mathbf{U}\mathbf{\Sigma}\mathbf{V}^T$ \citep{2002nreb.book.....V}. The kernels
can be found in the columns of $\mathbf{U}$ that correspond to zero-singular
values on the diagonal of $\mathbf{\Sigma}$. For the nuller architecture
described above, the rank of the matrix $\mathbf{A}$ is three, which means that
from the six outputs, three kernels can be assembled, a number that coincides
with the number of independent closure-phases one is expected to build with a
four-aperture interferometer.

For the special case where the phase shifting parameter of the mixing stage
$\theta=\pi/2$, this response matrix can be computed by hand, by plugging in
the first order approximation of the electric field described in Equation
\ref{eq:approx} to the right hand side of $\mathbf{M}$ and take the square
modulus: 

\begin{equation}
  \mathbf{A} = \frac{1}{4}
  \begin{bmatrix}
    1 & 1 & 1 & -1 &  0 &  0 \\
    1 & 1 & 1 & -1 &  0 &  0 \\
    1 & 1 & 1 &  0 &  0 & -1 \\
    1 & 1 & 1 &  0 &  0 & -1 \\
    1 & 1 & 1 &  0 & -1 &  0 \\
    1 & 1 & 1 &  0 & -1 &  0
  \end{bmatrix}.
  \label{eq:Amat}
\end{equation}

From here, it is easy to propose one possible kernel operator $\mathbf{K}$,
containing three linear combinations that erase all second instrumental phase
errors, by doing pairwise combinations of rows of $\mathbf{A}$:

\begin{equation}
  \mathbf{K} =
  \begin{bmatrix}
    1 & -1 &  0 &  0 &  0 &  0 \\
    0 &  0 &  1 & -1 &  0 &  0 \\
    0 &  0 &  0 &  0 &  1 & -1
  \end{bmatrix}.
  \label{eq:Kmat}
\end{equation}

\begin{figure}
  \includegraphics[width=\columnwidth]{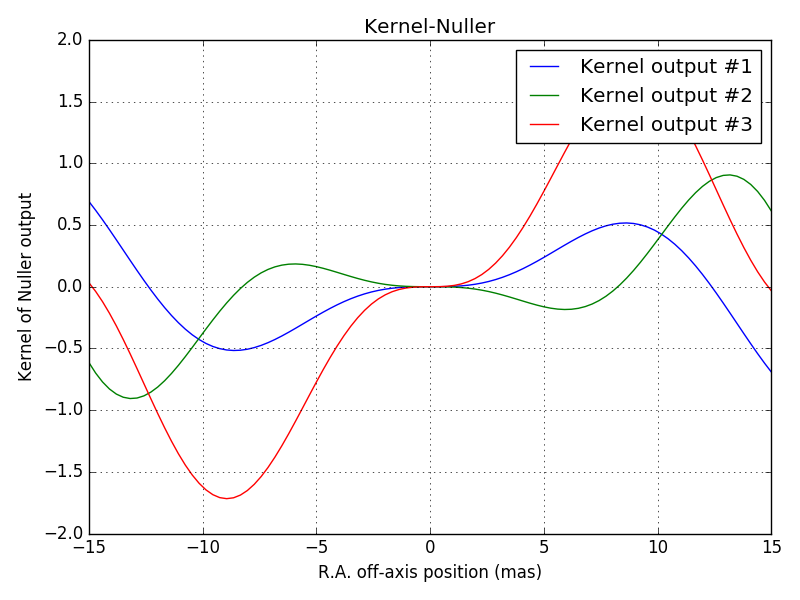}
  \caption{Evolution of the 3 kernels contained provided in Eq. \ref{eq:Kmat}
    as a function of R.A. offset relative to the reference null (in
    milli-arcsecond). The vertical unit of the plot is in multiples of the
    transmission of a single telescope.}
  \label{f:kernel_nuller_out_plots}
\end{figure}

The kernel outputs that are are primary observables are then:

\begin{equation}
\kvect = \mathbf{K} \cdot \outvect.
\end{equation}

Completing the description of the in-line interferometer introduced earlier,
Figure \ref{f:kernel_nuller_out_plots} shows how the three kernels $\kvect$ vary as a function of the position of the target, as it moves
across a $\pm$15 mas range of offset position relative to the null. The kernels
consisting of linear combinations of anti-symmetric response curves are also
anti-symmetric, just like closure- and kernel-phase.

Finally, we note that as our kernels are constructed as a linear combination of
output intensities, they have the same properties whether phase noise occurs
during an integration time or between integration times that are latter added
in post-processing. This is in contrast to nonlinear techniques such as nulling
self-calibration or closure-phase.

\section{Properties of a kernel-nuller for VLTI}
\label{sec:properties}

The high-contrast imaging properties of a nulling instrument, most notably the
general shape of the on-sky transmission map, will depend on the exact location
and size of the sub-apertures of the interferometer feeding light to the
recombiner. While the method outlined above is infrastructure-agnostic, we will
from now on examine at the special case of the VLTI, and describe the properties
of a instrument concept called VIKiNG, an acronym standing for the VLTI
Infrared Kernel NullinG instrument.

\subsection{Practical Implementation}
\label{sec:practical}

The direct detection of extrasolar planets with long baseline interferometry
points towards the use of the L-band (3.4 - 4.1 $\mu$m) where the blackbody
spectrum of forming planets is most likely to peak according to planet
formation models, and that of mature planets kept warm by the proximity of
their host star remains favorable. A viable practical implementation of both
nulling and sensing functions as shown in Figure~\ref{f:concept_modulation}
could rely on multi-mode interference (MMI) couplers made of Chalcogenide glass
(ChG) \citep{2013OExpr..2129927M} that provide good bandwidth at very close to
50/50 coupling and realistic fabrication tolerances
\citep{2017OExpr..25.3038K}. Both functions could be integrated into one single
photonic chip however fluctuations of the atmospheric thermal background will
require some form of modulation.  A bulk optics implementation, for instance
inspired by Figure~2 of \citet{2013PASP..125..951G} may also be possible.
  
One of the technological difficulties in designing space-based nulling
interferometers has been the ability to produce achromatic phase shifts and
50/50 couplers over large bandwidths. These problems do not go away in the
kernel-nulling approach proposed here, but we note that the requirements are
much more achievable for ground-based combiners aiming for the detection of
warm exoplanets. For example, phase shifts need only be significantly better
than the fringe tracking RMS, which is of order 100\,nm for the best current
fringe trackers. For a symmetrical physical nulling device such as the one
represented by our nulling matrix $\mathbf{N}$, input geometric phase shifts
between the inputs of $\pi$ are needed. A vacuum delay of 1.9\,$\mu$m achieves
a $\pi$ phase shift within 100nm for a 10\% bandwidth in the astronomical L'
band at $\sim$3.8\,$\mu$m, and simple first-order achromaticism with an
air-glass combination easily improves this by a factor of 10. The combination
of waveguide total length and core diameter can create similar achromaticity on
a chip.

\subsection{Nuller-output mapped on-sky}
\label{sec:remap}

\begin{figure*}
  \includegraphics[width=\textwidth]{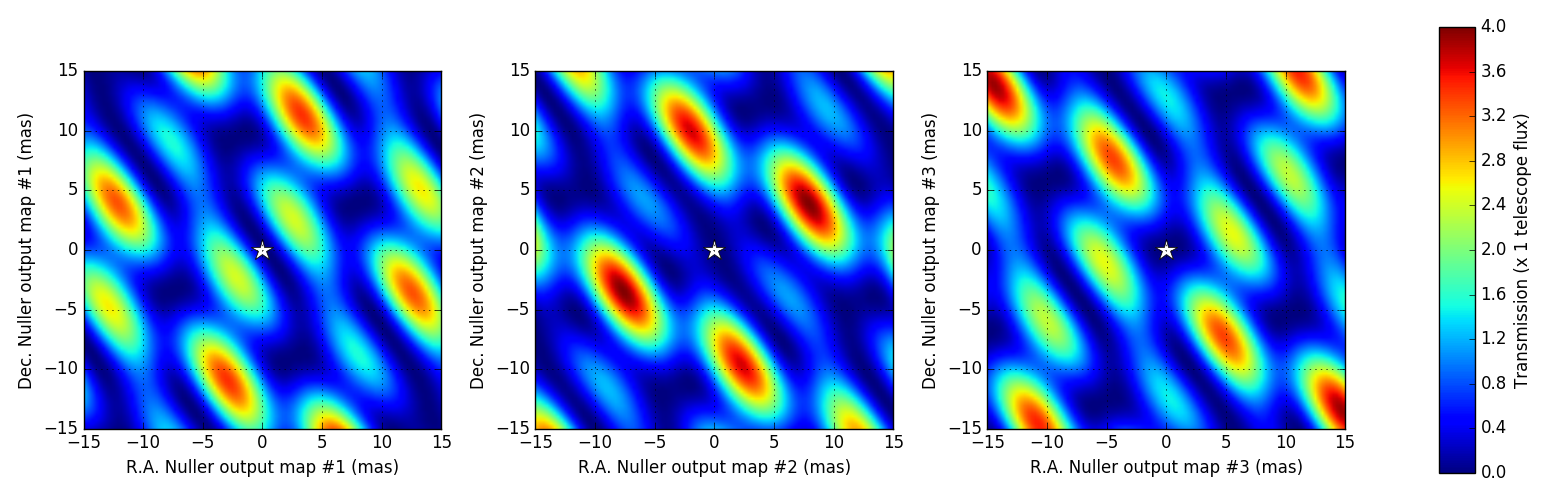}
  \caption{Transmission map for the three nulled outputs for a VLTI 4-UT
    aperture geometry over a $\pm$15 mas field of view. A five pointed star
    marks the location of the center of the field, where the rejection by the
    nuller is optimal. The three maps share the same colorbar, with a
    transmission that ranges from zero on the null to close to 100 \% (4 $F_T$)
    for a few places in the field whose positions are dictated by the geometry
    of the interferometric array.}
  \label{f:nuller_out_maps}
\end{figure*}

\begin{table}
  \caption{VLTI Unit Telescope East and North coordinates (in meters)}
  \label{tbl:uts}
  \centering
  \begin{tabular}{ l r r}
    \hline\hline
    Station & E & N \\
    \hline
    U1     &  -9.925 & -20.335 \\
    U2     &  14.887 &  30.502 \\
    U3     &  44.915 &  66.183 \\
    U4     & 103.306 &  44.999 \\
    \hline
\end{tabular}
\end{table}

Our study case will focus on the simultaneous use of the four 8-meter diameter
unit telescopes (UTs) of VLTI, pointing and cophased so as to observe a field
of view conveniently located exactly at zenith. The coordinates for these
stations, expressed in the reference system used to describe ESO's Paranal
observing facilities, are provided in Table \ref{tbl:uts}.  We start with the
nuller introduced in Section \ref{sec:design} and described by the unitary
matrix $\mathbf{N}$. It is used in the L-band at the wavelength $\lambda =
3.6\,\mu\mathrm{m}$.  For a snapshot observation, the field of view provided by
the intererometer is given by the shortest (46.6 meter) baseline size of the
array, corresponding to a $\sim$15 mas diameter.

In addition to the overall geometry of the array, the order by which the four
input beams are recombined into the nuller will impact the imaging properties
of the system. We will not attempt to optimize the nuller's performance by
re-ordering the input beams and will simply plug them in the order provided
by Table \ref{tbl:uts}. Figure \ref{f:nuller_out_maps} shows the resulting 2D
transmission maps for each of the three outputs of the nuller over a $\pm$15
mas field of view both in right ascension and declination. The transmission is
expressed in units of the flux collected by one aperture: $F_T$. As expected
from the analysis of the in-line array, the three maps are symmetric about the
origin: the transmission is zero on-axis, where the host-star would be
located. The geometric arrangement of the four apertures makes the nuller
observations, very much like any other interferometric observation,
non-uniformly sensitive over the field. Each output features a different
transmission profile that can peak up to close to 4 $F_T$ (corresponding to 100
\% transmission) that is more sensitive to the presence of a structure for
different parts of the field.

\begin{figure*}
  \includegraphics[width=\textwidth]{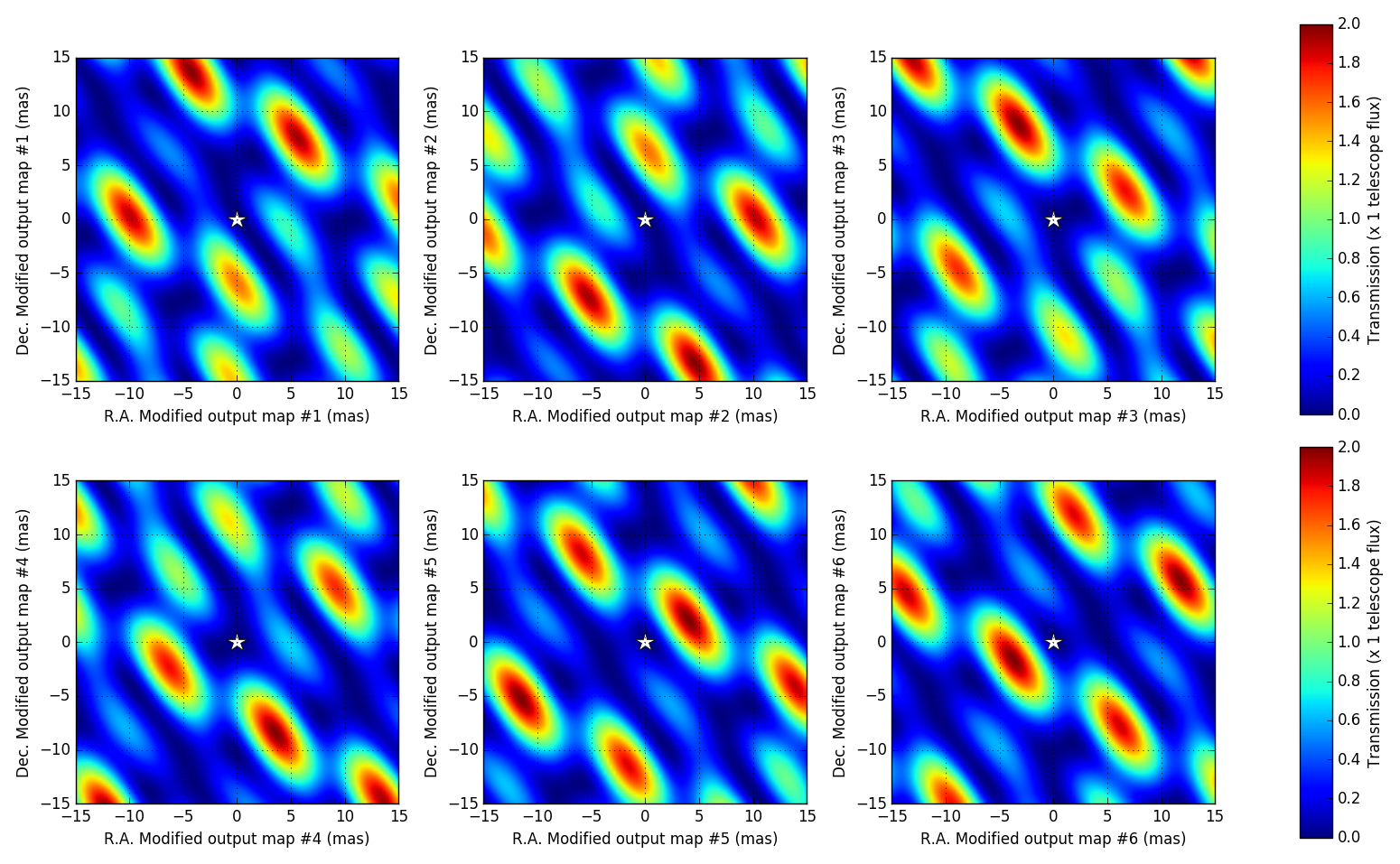}
  \caption{Transmission map for the six outputs of the modified nuller design
    for a VLTI 4-UT aperture geometry over a $\pm$15 mas field of view. A five
    pointed star marks the location of the center of the field, where the
    rejection by the nuller is optimal. All maps share the same colorbar, with
    a transmission that range from zero on the null to 50 \% of the total flux
    collected by the four apertures (2 $F_T$). Compared to the maps provided in
    Figure \ref{f:nuller_out_maps}, the amplitude of the colorscale was reduced
    by a factor of 2.}
  \label{f:raw_out_maps}
\end{figure*}

Figure \ref{f:raw_out_maps} shows how the six transmission maps of the modified
nuller vary over the same field of view. By doubling the number of outputs, one
expects the flux per output to be reduced by a factor of two: the colorscale of
the figure was therefore adjusted in consequence. The six new maps all have a significant 
anti-symmetric component about the center of the field, which means that in the absence
of perturbation, these six observables can better constrain the position of a
potential companion to an observed target.

\begin{figure}
  \includegraphics[width=\columnwidth]{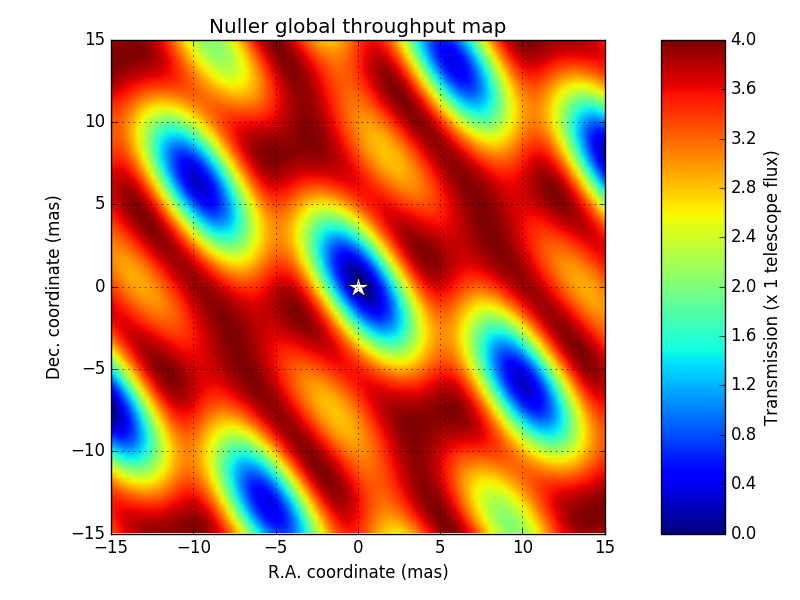}
  \caption{Map of the global throughput of the nuller, corresponding to the sum
    of the three maps provided in Fig. \ref{f:nuller_out_maps} or the six maps
    provided in Fig. \ref{f:raw_out_maps}.}
  \label{f:throughput}
\end{figure}

Note that the sum of these six new transmission maps for the modified nuller,
is identical to the sum of the three transmission maps of the original design:
in the absence of coupling losses between the nulling and the mixing stages,
the flux is simply redistributed amongst the different channels by the 3x6
combiner labeled $\mathbf{S}$ in Figure \ref{f:concept_modulation}. This global
transmission map is displayed in Figure \ref{f:throughput}: one can verify that
it is the complement to the on-axis fringe pattern produced by the VLTI 4-UT
array, rejected to the bright output of the nuller as illustrated in
Fig. \ref{f:concept_modulation}.

\begin{figure*}
  \includegraphics[width=\textwidth]{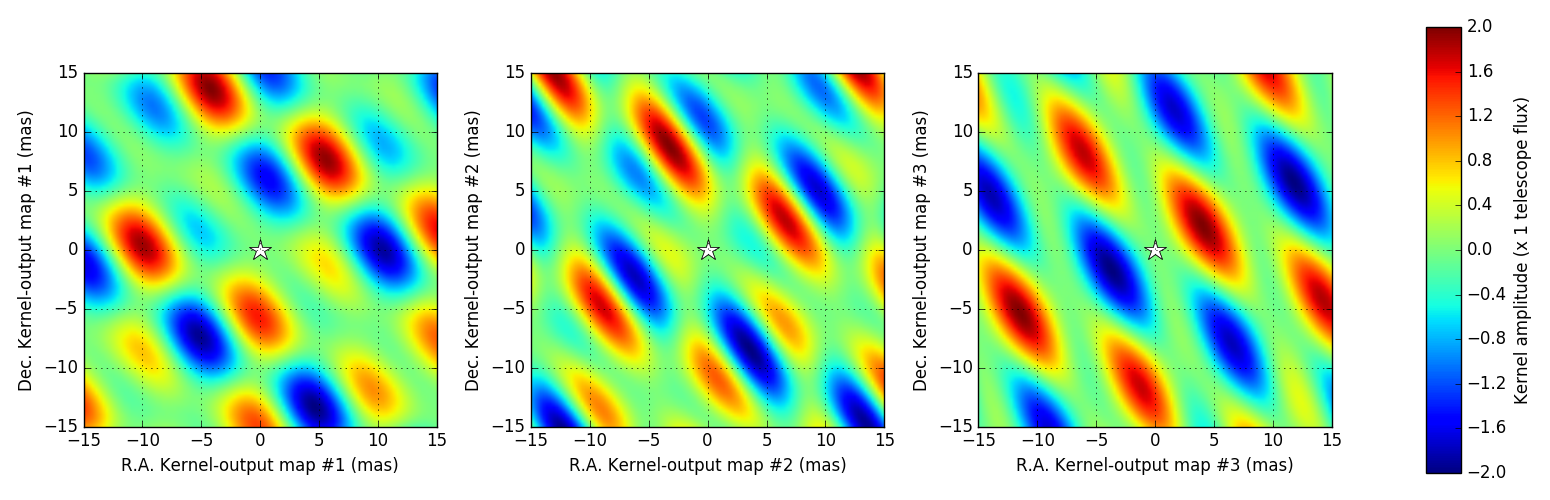}
  \caption{Evolution of the three kernel-outputs of the modified nuller
    architecture as a function of the position over a $\pm$15 mas field of
    view. Observe that all three maps are antisymmetric. The sign of the
    outputs can tell which side of the field of view a companion is.}
  \label{f:kernel_maps}
\end{figure*}

\subsection{Phase error robustness}
\label{sec:robustness}

We use the result of a series of simulated nulling observations that
demonstrate the interest of the modified architecture and its kernel. As
reminded by the different transmission maps used in the previous section, the
detectability of an off-axis structure by the nuller is not uniform over the
field of view. To ease our description, we will arbitrarily place a companion
with a contrast $c = 10^{-2}$ at the coordinates (+1.8, +4.8) mas in the system
used so far, where the sensitivity of the nuller $\mathbf{N}$ is near optimal
for the VLTI 4-UT (at zenith) configuration, as can be guessed by looking at
the global throughput map shown in Figure \ref{f:throughput}.

\begin{figure*}
  \includegraphics[width=\textwidth]{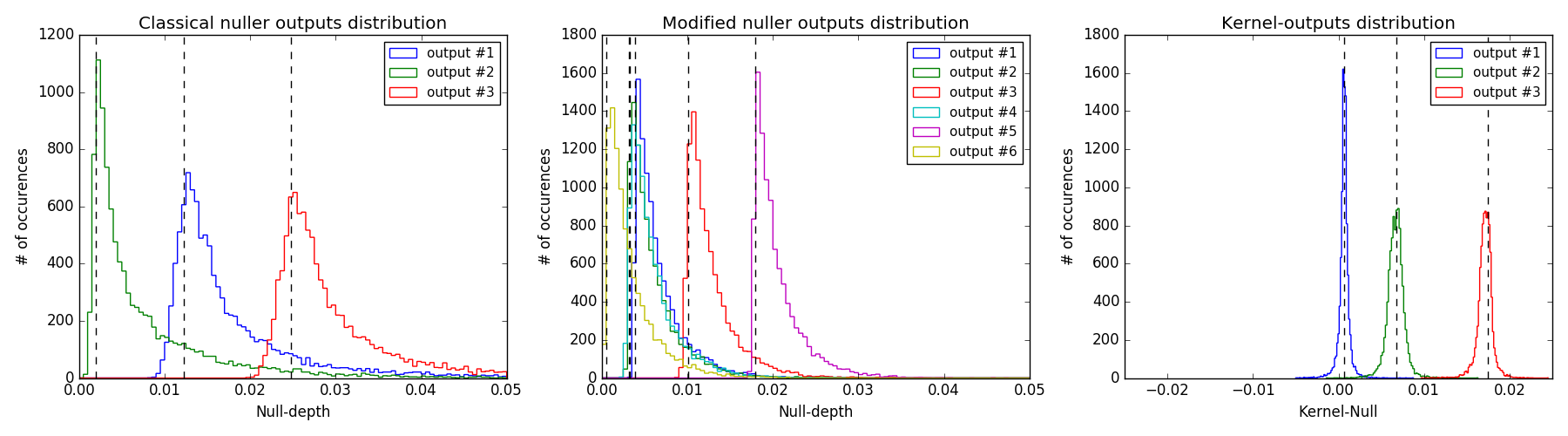}
  \caption{Distribution of the outputs of the nuller during the observation of
    a binary object (companion of contrast $c = 10^{-2}$ at (+4.8, +1.8) mas)
    in the presence of 50 nm RMS residual piston excursions drawn from a normal
    distribution. From left to right: the nuller alone, the nuller+sensor and
    the kernels. The dashed lines mark the expected location of the different
    nulls (and their kernels) in the absence of piston excursion.}
  \label{f:histo_50}
\end{figure*}

Figure \ref{f:histo_50} present the results of these simulations (a total of
$10^4$ acquisitions per simulation), in the presence of 50 nm residual piston
excursions. Each sub-figure features the histograms of outputs at the different
stages of the concept. The null-depth bin values quoted in these figures are in
units consistent with the transmission maps shown in Figures
\ref{f:nuller_out_maps} and \ref{f:raw_out_maps}: the null-depth bin for a
given output is proportional to the contrast of the companion, and multiplied
by the transmission of the nuller for these coordinates.

The expected transmission of the three dark outputs after the 4x4
nulling-coupler is $t = (1.22, 0.19, 2.47)$. For a $c= 10^{-2}$ contrast, one
expects, in the absence of residual piston errors, outputs of 0.0122, 0.0019
and 0.0247, marked in the left panel of Figure \ref{f:histo_50} by three
vertical dashed lines. In the presence of residual piston error, the
distribution of observed null-depth deviates from what should be a Dirac
distribution and evolves into the three plotted skewed distributions (see
\citet{2011ApJ...729..110H} for a formal model of this distribution).  A
real-world scenario with background and residual target shot-noise would
convolve this distribution with a Gaussian, complicating its interpretation.
The six outputs of the modified nuller design, including the mixing stage
provided by $\mathbf{S}$ are similarly distributed, and are equally affected by
the residual piston errors.

The raw nuller outputs spend very little time on the null and figuring out the
true value of the null requires careful modeling of this distribution. By
comparison, the kernel outputs, visible in the right panel of Figure
\ref{f:histo_50}, are well distributed and the statistics are relatively
straightforward. Consistent with the general results from
\citet{2013MNRAS.433.1718I}, the uncertainty in the kernel outputs is
proportional to the cube of the phase errors.

\subsection{Sensitivity}
\label{sec:sensitivity}

\begin{figure}
  \includegraphics[width=\columnwidth]{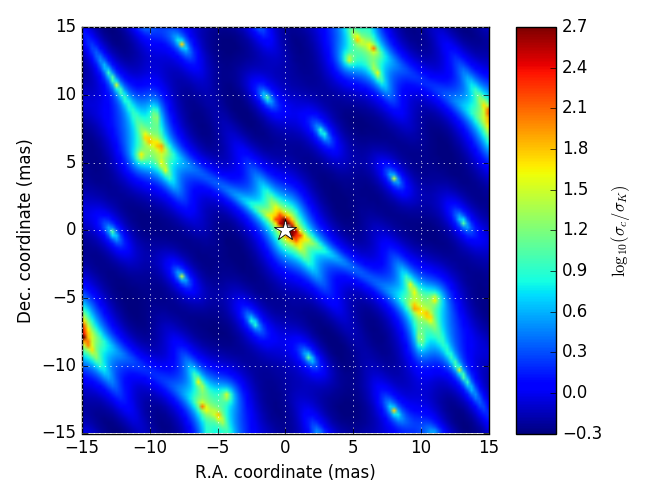}
  \caption{Map of the ratio between contrast uncertainty and kernel output
    uncertainty as a function of R.A. and Dec for the VLTI-4UT configuration.
    The map uses a logarithmic stretch, ranging from -0.3 ($\sigma_c = 0.5
    \times \sigma_k$) in the most favorable configurations to $\sim$3
    ($\sigma_c = 10^3 \times \sigma_k$) near the null. The median ratio is
    $\sigma_c = 0.8 \times \sigma_k$.}
  \label{f:contrast_error_gain}
\end{figure}

For a companion of known relative position $(\alpha, \delta)$, the contrast $c$
is the solution to:

\begin{equation}
  \knormvect = \mathbf{m} \times c,
\end{equation}

\noindent
where $\knormvect$ is a vector containing the measured three kernel-outputs
($\kvect$) normalised by the total flux (i.e. total including the starlight
output) and $\mathbf{m}$ a vector containing the values of the kernel
transmission maps (see Figure \ref{f:kernel_maps}) for the coordinates
$(\alpha, \delta)$. In the presence of uncertainties, the best estimate for $c$
is the least-square solution:

\begin{equation}
  c = (\mathbf{m}^T \cdot \knormvect) / (\mathbf{m}^T \cdot \mathbf{m}),
\end{equation}

\noindent
with associated uncertainty:

\begin{equation}
  \sigma_c = \frac{1}{|\mathbf{m}|} \sigma_k,
\end{equation}


\noindent
where $\sigma_y$ is the dispersion of the kernel-output estimate. The
$1/|\mathbf{m}|$ parameter scaling the two uncertainties depends on the
position of the companion in the field of view, as shown in Figure
\ref{f:contrast_error_gain}, and varies from $\sigma_c = 0.5 \times \sigma_k$
in the most favorable configurations to $\sigma_c = 10^3 \times \sigma_k$ near
the null, with a median ratio $\sigma_c = 0.8 \times \sigma_k$. 


There are four key fundamental sources of uncertainty which are added in
quadrature in forming the kernel-uncertainty $\sigma_k$: the fringe tracking
phase errors ($\sigma_{k,\varphi}$), the cross-term between the fringe-tracking
phase errors and intensity fluctuations on other telescopes ($\sigma_{k,
  I\varphi}$), the thermal background ($\sigma_{k,B}$) and the residual target
photon noise ($\sigma_{k,T}$). For the uncertainty derived from the
fringe-tracking phase, we can approximate the effect of many independent
wavefront realizations by modeling the fringe-tracker uncertainty power
spectrum as white up to a cutoff frequency $\Delta \nu_{\rm FT}$. This means
that there are $\nu_{\rm FT} \times \Delta T$ realizations of fringe tracker
errors, resulting in a contribution to the integrated kernel output uncertainty
$\sigma_{k}$ of:

\begin{align}
\sigma_{k, \varphi} &\approx \sigma_{\varphi}^3 \Delta \nu_{\rm FT}^{-1/2} \Delta T^{-1/2},
\end{align}

\noindent
This equation becomes accurate at the $\sim$10 \% level for $\sigma_{\varphi} <
0.3$, which we have verified through simulation.  Note that if the fringe
tracker does not average to zero phase offset, then this third order kernel
output uncertainty would not average to zero.  In practice, any systematic
offset in the fringe tracker zero point would have to be $\sim$10 times smaller
than $\sigma_{\varphi}$ in order to be insignificant for typical exposure times
and fringe tracker bandwidths. In the presence of precipitable water vapor,
this stringent requirement can be achieved in different ways such as carrying
out both nulling and fringe tracking at the same wavelength, fast
re-calibration of the nulling setpoints (faster than water vapor seeing), or by
dedicated control loops such as demonstrated with the KIN \citep{Koresko2006}
and the LBTI \citep{Defr2016}.

The cross-term between intensity fluctuations and piston tracking errors is a
second order term with a contribution to the kernel output uncertainty of:

\begin{align}
\sigma_{k, I\varphi} &\approx 2^{-1/2} \sigma_\varphi \sigma_I \Delta \nu_{\rm max} \Delta T^{-1/2},
\end{align}

\noindent
where $\sigma_I$ is the intensity fluctuation on each telescope, and $\Delta
\nu_{\rm max}$ is the maximum of the adaptive optics bandwidth (for fiber
injection) and the piston bandwidth. From \citet{2017A&A...604A.122J},
practical RMS coupling efficiency variations with an extreme adaptive optics
system can be of order 10\% at 1.55\,$\mu$m, which would correspond to
$\sim$2\% in the L' band. Coupling fluctuations are often much worse than this
for existing interferometers with adaptive optics, with one problem being
inadequate control of low-order modes.

The contribution of residual target photon shot noise is:

\begin{align}
\sigma_{k, T} &\approx \sigma_{\varphi}^2 F_T^{-1/2} \Delta T^{-1/2},
\end{align}

\noindent
where $F_T$ is the target flux in photons/s/telescope, and other symbols
are as before. The power of -1/2 is the combination of two terms: the $\sqrt{F_T}$ 
increase in the noise, and the scaling by $1/F_T$ in obtaining the normalised 
kernel outputs $\knormvect$ from the raw outputs $\kvect$.
The contribution of thermal background has a similar functional form for the same reason:

\begin{align}
\sigma_{k, B} &\approx F_B^{1/2} F_T^{-1} \Delta T^{-1/2}.
\end{align}

For observations in an L' filter (3.4 to 4.1 $\mu$m), we can write
\citep{2005PASP..117..421T} the target and background flux for a warm optics
temperature of 290\,K as:

\begin{align}
F_T &= 3.5 \times 10^{10} \eta_c \eta_w \left( \frac{D}{8\,{\rm m}} \right)^2 10^{-0.4 m_{L'}}{~\rm photons/s}. \\
F_B &=  A(T_w) \eta_c (1-\eta_w) {~\rm photons/s}. 
\end{align}

\begin{figure}
  \includegraphics[width=\columnwidth]{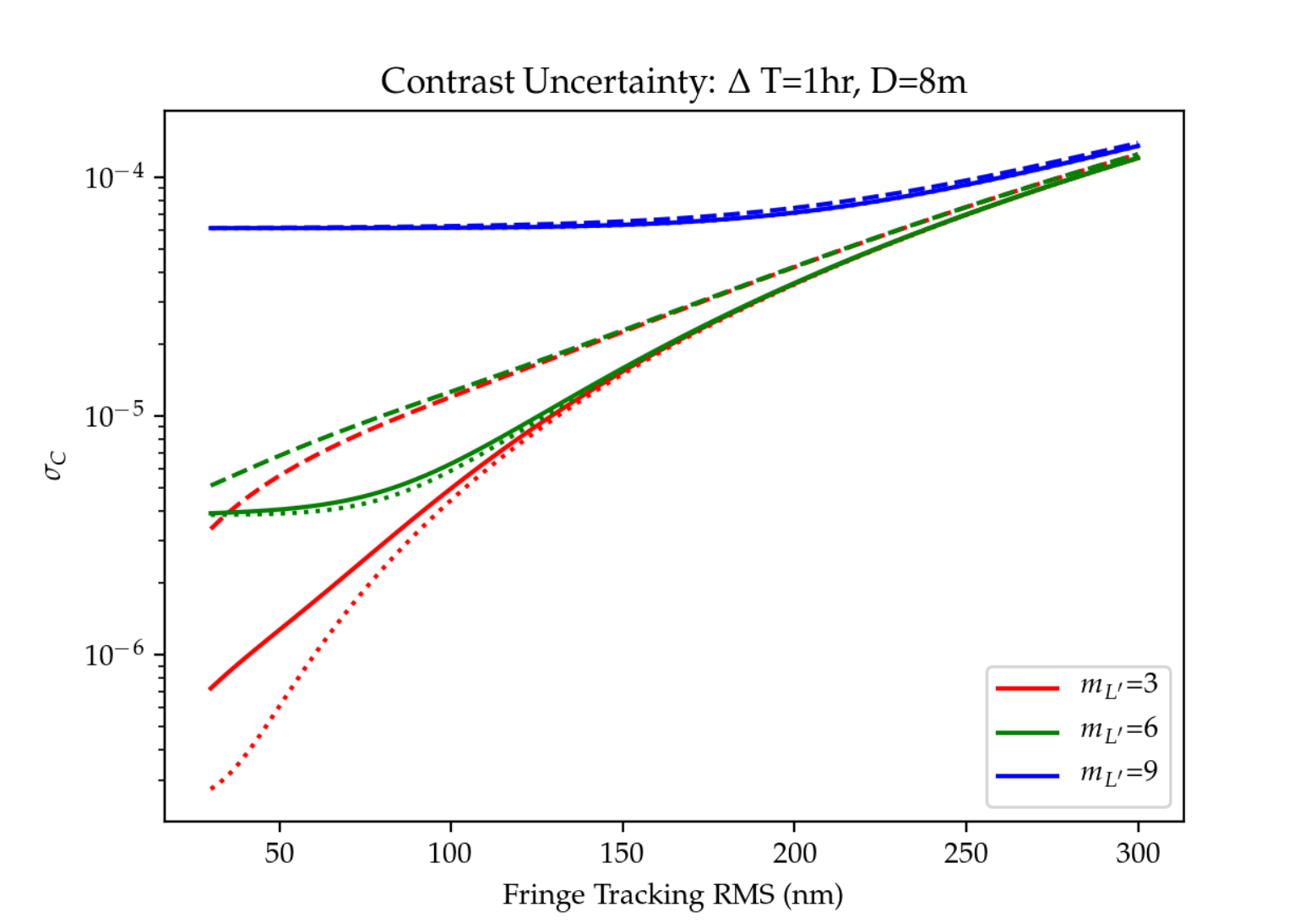}
  \caption{Contrast uncertainty (median over sky positions) as a function of
    fringe-tracker phase error, for different values of target
    magnitudes. Dotted line is for no intensity fluctuations, solid line for
    (realistic) 2\% RMS intensity fluctuations and dashed line for poor 10\%
    intensity fluctuations.  For high target fluxes, fringe tracker phase error
    dominates, and for low target fluxes, thermal background
    dominates. Residual target shot noise never dominates at an optics
    temperature of 290\,K.}
  \label{f:contrast_vs_phase_error}
\end{figure}

The background flux constant due to the warm telescope and interferometer
optics $A(T_w)$ per telescope is simply given by the Bose-Einstein distribution
applicable to photons applied to two polarisations and one spatial mode. Note
that this is the same as the Planck function in units of photons per unit
frequency applied to an \'etendue of $\lambda^2$.  This is $5.4 \times 10^7$
photons/s for 290\,K, and is generally given by:

\begin{align}
A(T_w) &= \frac{2 \Delta \nu}{\exp(h \nu/ k_b T_w) - 1},
\end{align}

\noindent
for a filter central frequency $\nu$ and bandwidth $\Delta \nu$.  With an
assumption of warm optics efficiency of $\eta_w=0.25$, and a cold optics
efficiency of $\eta=0.4$, the achievable contrast for 8\,m telescopes  is
shown in Figure~\ref{f:contrast_vs_phase_error}. These sensitivities are well
within the range needed to detect a range of transiting exoplanets, exoplanets
discovered by radial velocity and young, self-luminous exoplanets.

\subsection{The VIKiNG survey}
\label{sec:target_list}

The achievable contrast curves shown in Figure~\ref{f:contrast_vs_phase_error}
suggest that even with a conservative 150 nm RMS fringe tracking performance, a
contrast better than $c=10^{-5}$ can be achieved under realistic photometric
stability conditions for targets brighter than $M_L = 6$. A kernel-nulling
observing campain using the four VLTI UTs therefore presents a real potential
for the direct detection of nearby exoplanets discovered by radial velocity.
To support this claim, we used the information compiled in the Extrasolar
Planet Encyclopaedia database (\url+exoplanet.eu+) to select a sample of nearby
known extrasolar planet hosts that would make valuable targets for our VIKiNG
instrument concept, observing from VLTI at Paranal.

Selection criteria include a predicted contrast cutoff at $c=10^{-5}$, such
that a maximum total observing time of two hours per target makes it possible
to reach $SNR=5$, and an angular separation ranging from 5 to 15 mas. These
conservative inner and outer working angles respectively correspond to the
resolution of the longest and the shortest baselines for the assumed VLTI
configuration. The outer working angle could be extended by taking into account
the evolution of the (u,v) coverage over the observing time required to reach
the required SNR.

Assuming that these objects are at thermal equilibrium with their host star
allows to constrain a temperature (assuming albedo near zero, applicable to hot
Jupiters). We also assume an intermediate Neptune-like density ($1.64 g/cm^3$)
for all planets and use M $\sin i$ to put a lower constraint on the planet
radius. With temperature and radius estimates for both the star and the planet,
we can predict a contrast, while a angular separation estimate is simply given
by the ratio between the semi-major axis and the distance to the
system. Fourteen targets fit all of the requirements. They are listed along
with their predicted observational properties in Table \ref{tbl:targets}.

\begin{table}
  \caption{VIKiNG best targets}
  \label{tbl:targets}
  \centering
  \begin{tabular}{l c c}
    \hline\hline
    Planet & Separation & Contrast \\
    name & (mas) & ($\log_{10}{c}$) \\
    \hline
    GJ 86 A b    & 10 & -4.03 \\
    BD+20 2457 b & 7  & -4.56 \\
    HD 110014 c  & 7  & -4.56 \\
    11 Com b     & 12 & -4.59 \\
    ksi Aql b    & 9  & -4.61 \\
    61 Vir b     & 6  & -4.67 \\
    HIP 105854 b & 10 & -4.68 \\
    HIP 107773 b & 7  & -4.75 \\
    HD 74156 b   & 5  & -4.84 \\
    mu Ara c     & 6  & -4.86 \\
    HD 168443 b  & 8  & -4.87 \\
    HIP 67851 b  & 7  & -4.91 \\
    HD 69830 b   & 6  & -4.94 \\
    HD 16417 b   & 5  & -4.99 \\
    \hline
\end{tabular}
\end{table}

The size of this sample doubles \citep{2018arXiv180707467D}, if one assumes a
tighter inner working angle of 1 mas ($0.25 \times \lambda/B$), which brings in
potentially warmer planets with more favorable contrasts. A more detailed
characterization of the true VIKiNG discovery and characterization potential is
beyond the scope of this paper that only aims at introducing a new instrument
concept. It will likely be the object of future work, to be carried out in the
context of the Hi-5 initiative \citep{2018arXiv180104148D}.

\section{Discussion}
\label{sec:disc}

\citet{2014MNRAS.439.4018L} proposed a different architecture concept for an
interferometric nuller able to produce closure-phase measurements of nulled
outputs. In the framework of this paper, the imaginary components of all three 
visibilities from those ABCD combiners are kernel outputs, and the imaginary 
component of the triple product simulated in that paper is just one
of three robust observables. However, in the critical background-limited regime, 
using all three kernel-outputs in the combiner of \citet{2014MNRAS.439.4018L}
would require an exposure time 6 times larger than the architecture presented 
here (Figure~\ref{f:concept_modulation}). We have also argued here that 
a linear combination of outputs is adequate for high contrast imaging, without the 
need for the nonlinear operations of creating triple products or 
computing closure-phase.

It should also be observed that the methodology outlined earlier can also be
applied to show that, the nulling observations are rendered robust against
inter-beam intensity fluctuations, due either to high-altitude atmospheric
turbulence (scintillation) or to intra-beam high-order wavefront aberrations
that result in coupling losses.  The null is also a quadratic function of these
intensity fluctuations \citep{2012ApJ...748...55S}.  While sensitive to
photometric unbalance, the behavior of the nuller remains insensitive to global
fluctuations of the source brightness. Like for the piston, with the flux of
one sub-aperture taken reference, there are only six second-order relative
perturbations terms that will impact the nuller's outputs.  The impact of these
fluctuations can be modeled using the framework outlined for the phase,
substituting in Equation~\ref{eq:approx}, a real phase term $\varphi_k$ for an
imaginary term, that results in an electric field with a modulus that deviates
from unity.  The structure of the resulting response matrix $\mathbf{A}$ is
identical to the one for the phase: the same kernel matrix $\mathbf{K}$ will
therefore simultaneously render the observable quantities robust against piston
excursions and small amplitude photometric fluctuations: the uncertainty in the
kernel-outputs is also proportional to the cube of the input complex amplitude
fluctuations, so that even 10\% intensity fluctuations on the inputs would
translate into errors smaller than $10^{-3}$ on the kernel-outputs.

We have however reported that the coupling between fringe tracking errors and
intensity fluctuations does contribute to the error budget as highlighted by
\citet{Lay2004}. Our simulations suggest that under realistic (2 \%) intensity
fluctuations, these cross-terms do not significantly degrade our predicted
performance.

\section{Conclusion}
\label{sec:conclu}

High-contrast imaging solutions thus far implemented, either in the context of
single-telescope coronagraphy or multi-aperture interferometry, have been
conceived on the premiss of the optical subtraction of the static diffraction
pattern produced by a stable on-axis source. The effective high-contrast
detection potential of such static solutions is, in practice, severely limited
by the least amount of wavefront perturbation that quickly drives otherwise
near-ideal solutions away from their high-contrast reference point.

Drawing on the idea of kernel, here applied to the outputs of an
interferometric nuller, we have described how the design of an otherwise plain
four input beam interferometric nuller can be modified to take into account,
the possibility of self-calibration. The result is a concept that, assuming
good but no longer ideal observing conditions, becomes robust against residual
wavefront aberrations (as well as photometric fluctuations), with errors
dominated by third order input phase and intensity errors.

Kernel-nulling interferometry is a powerful idea: the architecture and method
outlined in this paper make it possible to simultaneously benefit from the
high-contrast boost provided by the nuller while keeping the ability to sense
the otherwise degenerate effect of ever-changing observing conditions, so as to
build observable quantities that are robust against those spurious
effects. Similarly to closure-phase, our kernel-nulled outputs also break the
symmetry degeneracy of a classical nuller's output: the sign of the different
kernels constrains which side of the field of view any asymmetric structure
lies. Preliminary simulations suggest that under reasonable observing
conditions, our VIKiNG instrument concept, using the four UTs of the VLTI
infrastructure, could directly detect a dozen nearby planets discovered by
radial velocity surveys, in less than two hours spent per target.

Note that with only four input beams, the special case described in
this paper features a small number of possible covariance terms to keep track
of. Future work will attempt to answer the questions: ``Can the approach be
further generalized and applied to situations where a large number of degrees
of freedom are available?''  and ``How can a coronagraph be modified in order
to benefit from similar properties?''

The proposed concept is of course not restricted to ground-based
interferometry. The robustness boost brought by the concept of kernel-output
reduces the otherwise demanding technological requirements on a space borne
interferometer tasked with the direct detection of higher contrast ($10^{-10}$)
Earth-like extrasolar planets. It would be valuable to brushup the original
designs for the Darwin and TPF-I concept missions and see what a revised
kernel-nulling architecture can bring.

\begin{acknowledgements}
This project has received funding from the European Research Council (ERC)
under the European Union's Horizon 2020 research and innovation program (grant
agreement CoG \# 683029). MI was supported by the Australian Research Council
fellowship FT130100235. The collaboration represented by this paper was
supported by the Stromlo distinguished visitor program. The work benefited
greatly from discussions with Harry-Dean Kenchington Goldsmith and Stephen
Madden, as well as discussions at the Hi-5 meeting in Liege in 2017.  The
referee's detailed comments helped to greatly improve the manuscript.
\end{acknowledgements}

\bibliographystyle{aa}

\end{document}